\input harvmac
\input epsf

%
%

\nref
\DasI{ A. Das and S. Roy, ``On M-Theory and the Symmetries of
Type II String Effective Actions," hep-th/9605073. }

\nref
\DavisI{ K. Davis, ``M-Theory and String-String Duality,"
hep-th/9601102. }

\nref
\DavisII{ K. Davis, ``Enhanced Gauge Symmetry in M-Theory,"
hep-th/9608113. }

\nref
\DuffI{ M.J. Duff, R.R. Khuri, and J.X. Lu, ``String Solitons,"
hep-th/9412184; Physics Reports {\bf 259} ( 1995 ) 213-326 }

\nref
\GinspargI{ P. Ginsparg, ``On Torodial Compactification of Heterotic
Superstrings," Physical Review D Volume 35, Number 2 ( 648 - 654 ) }

\nref
\HullI{ C. Hull and P. Townsend ``Unity of Superstring Dualities,"
hep-th/9410167; Nuclear Physics B {\bf 438} (1995) 109-137.  }

\nref
\SchwarzI{ J. Schwarz, ``An SL(2,Z ) Multiplet of Type IIB
Superstrings," hep-th/9508143. }

\nref
\SeibergI{M. Dine, P. Huet and N. Seiberg, ``Large and Small
Radius in String Theory," Nuclear Physics B {\bf 322} (1989) 301-316 }

\nref
\SenI{A. Sen, ``String String Duality Conjecture in Six Dimensions
and Charged Solitonic Strings," hep-th/9504027, TIFR-TH-95-16;
Nuclear Physics B {\bf 450} (1995) 103-114 }

\nref
\WittenI{E. Witten, ``String Theory Dynamics in Various
Dimensions," hep-th/9503124; IASSNS-HEP-95-18;
Nuclear Physics B {\bf 443} (1995) 85 - 126  }

\nref
\WittenII{ E. Witten, ``Some Comments on String Dynamics,"
hep-th/9507121 }

\nref
\WittenIII{P. Horava and E. Witten, ``Heterotic and Type I String
Dynamics from Eleven Dimensions," hep-th/9510209;
IASSNS-HEP-95-86; PUPT-1571 }

\nref
\WittenIV{ E. Witten, ``Five-Branes and M-Theory on an Orbifold,"
hep-th/9512219; IASSNS-HEP-96-01}

\nref
\WittenV{ E. Witten, ``Phase Transitions in M-Theory and
F-Theory," hep-th/9603150; IASSNS-HEP-96-26 }

%
%

\Title{
	\vbox{
		\baselineskip12pt
		\hbox{ RU 96-XX }
		\hbox{ hep-th/9609005 }
		\hbox{ Rutgers Theory }
		}
	}
	{
	\vbox{
		\centerline{ A NOTE ON }
		\vskip 8pt
		\centerline{ TENSIONLESS STRINGS }
		\vskip 8pt
		\centerline{ IN }
		\vskip 8pt
		\centerline{ M-THEORY }
		}
	}

\centerline{ Kelly Jay Davis }
\bigskip
\centerline{Rutgers University}
\centerline{Serin Physics Laboratory}
\centerline{Piscataway, NJ 08855}

%
%

\vskip .3in
\centerline{ABSTRACT}
In this article we examine the appearance of tensionless strings in
M-Theory. We subsequently interpret these tensionless
strings in a String Theory context.  In particular,  we examine
tensionless strings appearing in M-Theory on $S^{1}$, M-Theory
on $S^{1} / {\bf Z}_{2}$, and M-Theory on $T^{2}$; we then interpret
the appearance of such strings in a String Theory context. Then we
reverse this process and examine the appearance of some
tensionless strings in String Theory. Subsequently we interpret
these tensionless strings in a M-Theory context.
\Date{8/20/96}

%
%

\newsec{
		Introduction
	}

The theory formerly known as string theory has undergone quite
a revolution as of late. In the past year and a half the five string
theories have all been related so as to form a single String Theory.
Older results \SeibergI\ relating the two Type II string theories
and the two Heterotic string theories \GinspargI\ have been
combined with newer results \SenI\ relating  the Type II and
Heterotic string theories so as to relate the Type IIA, Type IIB,
$E_{8} \times  E_{8}$ Heterotic, and $SO(32)$ Heterotic string
theories.  Also,  as of late, the  $SO(32)$  Heterotic string theory
has been related to  the Type I string theory \WittenIII\ thus
completing the ``loop" and relating all five consistent string theories
and forming a single String Theory. Furthermore, there has also
been progress on yet  another  front,  relating  String  Theory  to
M-Theory \WittenI \WittenIII,  an eleven-dimensional  theory
containing two-branes and five-branes \DuffI \WittenIV .  In
relating M-Theory to String Theory one finds that basically all
properties of String Theory may be derived from M-Theory.
One may derive these String Theory properties by
relating M-Theory to any given string theory, then
relating the given string theory to all the other string theories
by way of the various ``string-string dualities." In this way one
may derive various properties of String Theory from M-Theory.

In this article we will employ some of the relations
between M-Theory and String Theory to study the appearance
of tensionless strings in M-Theory. We will subsequently
interpret the appearance of these tensionless strings in a String
Theory context. Also, we will reverse this process and examine the
appearance of tensionless strings in various string theories and
interpret these tensionless strings in terms of M-Theory. The
purpose of this study is to try and better understand so-called
``phenomena of the second kind" \WittenV\ in which a
$p$-brane becomes tensionless\foot{ Note, ``phenomena of
the second kind" in \WittenV\ referred only to strings becoming
tensionless; however, the ``spirit" of the term is maintained if
we extend this to $p$-branes becoming tensionless. }.

In particular, we will first consider the appearance of a tensionless
string in M-Theory on $S^{1}$. M-Theory on $S^{1}$ is equivalent
to the Type IIA string theory \WittenI. Obviously the Type IIA string
theory possess a one-brane. From a M-Theory point-of-view this
one-brane arises from a M-Theory two-brane wrapping about
$S^{1}$. The tension of the resultant one-brane is given by
$T_{2,M}R_{11}$ \WittenII, where $T_{2,M}$ is the tension
of the two-brane and $R_{11}$ the $S^{1}$ radius. Hence, such a
one-brane becomes tensionless as $R_{11} \rightarrow 0$. We
will examine this limit from a Type IIA perspective. Also,
we will reverse this process and consider a tensionless string
appearing in the Type IIA string theory then interpret this
tensionless string  from a M-Theory perspective. After this
we will  move on to M-Theory on $S^{1} / {\bf Z}_{2}$.

Witten and Horava proved \WittenIII\ that M-Theory on
$S^{1} / {\bf Z}_{2}$ is equivalent to the $E_{8} \times E_{8}$
Heterotic string theory. The $E_{8} \times E_{8}$ Heterotic
string theory obviously possess a one-brane. From a
M-Theory point-of-view this one-brane
arises from a M-Theory two-brane wrapping around $S^{1} /
{\bf Z}_{2}$. The tension of the resultant one-brane is given
by $T_{2,M}R_{11}$, where here $R_{11}$ is the radius of
$S^{1} / {\bf Z}_{2}$.  Again, in this case the string becomes
tensionless in the limit $T_{2,M}R_{11} \rightarrow 0$. We will
examine this limit from a Heterotic perspective. After this
we reverse this situation and consider a tensionless string
in the $E_{8} \times E_{8}$ Heterotic string theory and
interpret it from a M-Theory perspective. Finally, we
consider M-Theory on $T^{2}$.

M-Theory on $T^{2}$, as was shown by Witten \WittenI, is
equivalent to the Type IIA string theory on $S^{1}$. The
Type IIA string theory on $S^{1}$ obviously possess a
one-brane; in fact, it possess two different one-branes. The
first one-brane is the standard one-brane of the Type IIA
string theory and the second one-brane arises from a
two-brane of ten-dimensional Type IIA string theory
wrapping about the $S^{1}$ factor. From a M-Theory
perspective these one-branes have a similar origin.
The M-Theory two-brane can wrap about either
one of the $S^{1}$ factors of $T^{2} = S^{1} \times S^{1}$.
Wrapping about the first $S^{1}$ leads to one of the
one-branes and wrapping about the second $S^{1}$ leads to the
other one-brane. If we denote the radius of the first $S^{1}$
as $R_{10}$ and the radius of the second $S^{1}$ as $R_{11}$,
then the tension of the first one-brane is $T_{2,M}R_{10}$
and the tension of the second one-brane is $T_{2,M}R_{11}$.
So, in this situation we can obtain a tensionless one-brane
by taking either $R_{10}$ or $R_{11}$ to zero. We will
interpret these limits in terms of a Type II string theory
on $S^{1}$. We then conclude with some general remarks
on the various limits studied in this article.

%
%

\newsec{
		Tensionless Strings : M-Theory on $S^{1}$
	}

In this section we will examine a tensionless string appearing in
M-Theory on $S^{1}$ from a Type IIA perspective. We will then
examine a tensionless string in the Type IIA string theory from
a M-Theory perspective. Let us start by examining the tensionless
string in M-Theory on $S^{1}$ from a Type IIA perspective.

%
%

\subsec{
		Tensionless Strings in M-Theory on $S^{1}$
	}

In this subsection we will examine a tensionless string appearing
in M-Theory on $S^{1}$ from a Type IIA perspective. Let us now
start this examination.

M-Theory on $S^{1}$ is equivalent to the Type IIA string theory
\WittenI. The Type IIA string theory obviously possess a one-brane.
From a M-Theory perspective this one-brane arises from a
two-brane wrapping about $S^{1}$. If we denote the radius of
the $S^{1}$ as measured in the M-Theory metric as $R_{11}$
and the tension of the two-brane as measured in the M-Theory
metric as $T_{2,M}$, then the tension of this one-brane $T_{1,M}$
in the M-Theory metric is
\eqn
\OneBraneTensionMTheoryMetric{
	T_{1,M} = T_{2,M} R_{11}.
}
\noindent So, one can see that if we wish the tension of the
two-brane $T_{2,M}$ to remain constant, then taking the tension
of the one-brane to zero entails taking the limit $R_{11} \rightarrow
0$. We will consider only this limit and not the limit in which
$T_{2,M} \rightarrow 0$ as we are trying to understand the
consequences of only a tensionless string and are not
concerned at this point with tensionless two-branes. Also, we
will assume that the M-Theory ten-metric $g_{10,M}$ behaves as
$g_{10,M} \rightarrow g_{10,M}$ in this limit. So, in this limit the
M-Theory target space takes the form seen in Figure 1A.

\centerline{
	\epsfbox{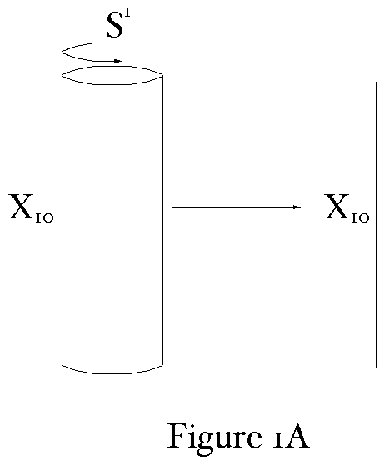}
}

\noindent where M-Theory is on $X_{10} \times S^{1}$ and $X_{10}$
is an arbitrary ten-manifold. Now, let us consider this from a
Type IIA perspective.

Employing the fact that M-Theory on $S^{1}$ is equivalent to
the Type IIA string theory one can obtain various relations
between the quantities defining the M-Theory compactification
on $S^{1}$ and the parameters of the Type IIA string theory.
In particular, one finds \DavisI \WittenI\ that the Type IIA
ten-dimensional coupling constant
$\lambda_{10,IIA}$ is related to the $S^{1}$ radius
as measured in the M-Theory metric $R_{11}$,
\eqn
\TypeIIACouplingRRelation{
	\lambda_{10,IIA} = R^{3/2}_{11}.
}
\noindent Furthermore, \DavisI \WittenI\ one finds that the
ten-dimensional M-Theory metric $g_{10,M}$ is related to the
ten-dimensional Type IIA metric $G_{10,IIA}$ as
\eqn
\TypeIIAMetricMTheoryMetricRelation{
	G_{10,IIA} = R_{11} g_{10,M}.
}
\noindent So, from this point-of-view one can see that the
Type IIA string theory is singular in the limit $R_{11}
\rightarrow 0$.

The metric of the ten-dimensional manifold, from a Type
IIA string theory perspective, ``vanishes." This is a result
of the fact that we require the M-Theory metric $g_{10,M}$
of the ten-manifold $X_{10}$ to be invariant in the limit
$R_{11} \rightarrow 0$. From this Type IIA perspective the
Type IIA target space takes the rather unfortunate singular
form seen in Figure 1B.

\centerline{
	\epsfbox{ 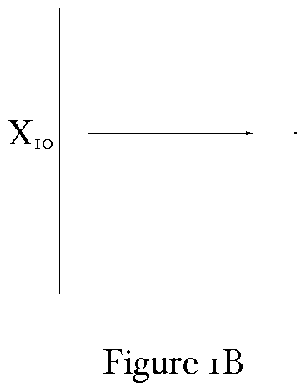 }
}

\noindent Now, let us consider how in the limit $R_{11}
\rightarrow 0$ the tensions of the M-Theory two-brane
and wrapped one-brane appear in the Type IIA string
theory.

Generically, a $p$-brane has a tension $T_{p}$ with dimension
$( $ length $)^{ -(p+1) }$. So, in particular, as the tension of a
$p$-brane is a dimensionful quantity, it depends upon the
metric in which it is measured. As the M-Theory ten-metric
and the Type IIA ten-metric are scaled relative to one-another
\TypeIIAMetricMTheoryMetricRelation, the tension of
any M-Theory $p$-brane is scaled relative to the tension
of its Type IIA counterpart. Hence, if we write the tension
of the two-brane as measured in the M-Theory ten-metric 
as $T_{2,M}$, then the tension of the two-brane $T_{2,IIA}$
in the Type IIA ten-metric is
\eqn
\TypeIIATwoBraneTension{
	T_{2,IIA} = R^{-3/2}_{11} T_{2,M},
}
\noindent where the equality follows from
\TypeIIAMetricMTheoryMetricRelation.
Employing similar logic we may calculate the
tension of the one-brane in the Type IIA string theory. First
though, remember that the one-brane arises from wrapping
the M-Theory two-brane around $S^{1}$ and hence has a
tension $T_{1,M}$ given by $T_{1,M} = T_{2,M} R_{11}$ in the
M-Theory metric \OneBraneTensionMTheoryMetric.
Measuring this tension in the Type IIA metric one
finds a tension $T_{1,IIA}$ given by,
\eqn
\TypeIIAOneBraneTension{
	T_{1,IIA} = R^{-1}_{11} T_{1,M} = T_{2,M},
}
\noindent where the first equality follows from
\TypeIIAMetricMTheoryMetricRelation\ and
the second from \OneBraneTensionMTheoryMetric.

Hence, in the limit $R_{11} \rightarrow 0$, the limit
in which M-Theory obtains a tensionless string
and a two-brane of finite tension, the Type IIA theory
does not possess a tensionless string by way of equation
\TypeIIAOneBraneTension, but the Type IIA theory
does possess a two-brane of infinite tension, in accord
with equation \TypeIIATwoBraneTension. In addition,
according to equation \TypeIIACouplingRRelation, the
Type IIA theory is weakly coupled in this region. As to
what this all ``means," I do not know. 

However, what is certain is that for an arbitrary $X_{10}$
we do not have a handle on the theory in this particular
limit.  From a Type IIA point-of-view,
spacetime is becoming singular and in addition
$T_{2,IIA} \rightarrow \infty$ . One can see this in
Figure 1B or equivalently in equation
\TypeIIAMetricMTheoryMetricRelation\ as $R_{11}
\rightarrow 0$ and $g_{10,M} \rightarrow g_{10,M}$.
So, for general $X_{10}$ we really do not know what is going
on from a Type IIA perspective as we ``generically" have no
perturbative framework in which to work. However, in
various special cases one may at least ``resolve" the
singular ten-manifold $X_{10}$ from a Type IIA
perspective. A particular example of this is if $X_{10} =
T^{10}$. In this case one could perform a T-Duality
transformation on each of the ten radii and obtain a
Type IIA theory on ${\bf R}^{10}$. However, this is
only a special case, and our understanding of this situation
for a general $X_{10}$ is minimal. So, let us move on and
try to understand the tensionless strings of the Type
IIA string theory from a M-Theory point-of-view.

%
%

\subsec{
		Tensionless Strings in Type IIA String Theory
	}

As we found in the previous subsection, the limit in which
the one-brane of M-Theory on $S^{1}$ becomes tensionless
is ``generically" rather ill behaved. In this section we will reverse
the process of last section. We will consider the appearance of a
tensionless string in the Type IIA string theory, then we will
interpret the appearance of this string in a M-Theory context.
Most of the ``leg-work" for this investigation was done in the
previous subsection; so, we will rely heavily upon the previous
subsection in our calculations.

Let us consider the limit in which the tension of the
Type IIA one-brane vanishes $T_{1,IIA} \rightarrow 0$.
Furthermore, let us also assume that $G_{10,IIA}$ and
$T_{2,IIA}$ do not vary in this limit, $G_{10,IIA} \rightarrow
G_{10,IIA}$ and $T_{2,IIA} \rightarrow T_{2,IIA}$. Graphically,
this is represented in Figure 2A.

\centerline{
	\epsfbox{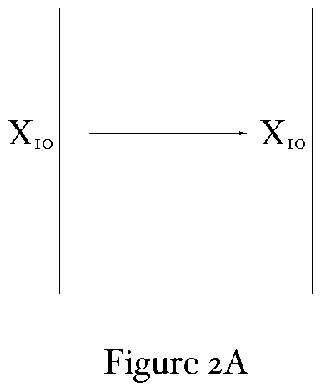}
}

As $T_{1,IIA} = T_{2,M}$ according to equation
\TypeIIAOneBraneTension, one can see that in the limit
$T_{1,IIA} \rightarrow 0$ the M-Theory two-brane tension
$T_{2,M}$ goes to zero. Similarly, as $T_{2,IIA} = R^{-3/2}_{11}
T_{2,M}$ in accord with equation \TypeIIATwoBraneTension,
$R_{11} \rightarrow  0$ in this limit due to the fact that $T_{2,M}
\rightarrow 0$ and $T_{2,IIA} \rightarrow T_{2,IIA}$.
Furthermore, as the Type IIA ten-metric and the M-Theory
ten-metric are related as in equation
\TypeIIAMetricMTheoryMetricRelation, the limit implies that
$g_{10,M}$ becomes ``flat." Graphically, this is depicted in
Figure 2B

\centerline{
	\epsfbox{ 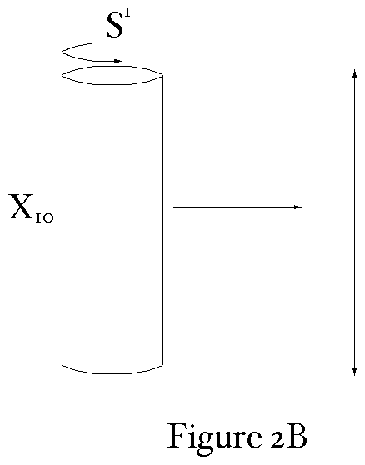 }
}
\noindent where the arrowheads denote the fact that the
ten-manifold from a M-Theory point-of-view is infinitely
``stretched."

In this limit, for general $X_{10}$, M-Theory is singular. One can
see this relatively easily from Figure 2B or equivalently from the
fact that $R_{11} \rightarrow 0$. But, for special $X_{10}$
of the form $X_{10} = X_{9} \times S^{1}$, where $X_{9}$ is an
arbitrary nine-manifold, one can perform a T-Duality transformation
on the collapsing $S^{1}$ \SeibergI\ so as to yield a theory with a
non-singular spacetime. We will examine this interesting property
properly in section four. So, we will not discuss it further here.

%
%

\newsec{
		Tensionless Strings : M-Theory on $S^{1} / {\bf Z}_{2}$
	}

In this section we will examine a tensionless string appearing in
M-Theory on $S^{1} / {\bf Z}_{2}$ from an $E_{8} \times E_{8}$
Heterotic string theory perspective. We will then examine a
tensionless string in the $E_{8} \times E_{8}$ Heterotic string
theory  from a M-Theory perspective. Let us start, however, by
examining the tensionless string in M-Theory on $S^{1} / {\bf Z}_{2}$
from an $E_{8} \times E_{8}$ Heterotic string theory perspective.

%
%

\subsec{
		Tensionless Strings in M-Theory on $S^{1} / {\bf Z}_{2}$
	}

In this subsection we will examine a tensionless string in M-Theory
on $S^{1} / {\bf Z}_{2}$ from an $E_{8} \times E_{8}$ Heterotic string
theory perspective. This examination, as we will find, is very similar
to the Type IIA examination in the previous section. However, we
will also find several critical differences. Let us now proceed with this
examination.

M-Theory on $S^{1} / {\bf Z}_{2}$, as was proven by Witten and Horava
\WittenIII, is equivalent to the $E_{8} \times E_{8}$ Heterotic string
theory. If we denote the radius of the $S^{1} / {\bf Z}_{2}$ factor as
measured in the M-Theory metric by $R_{11}$, then one finds
\DavisII \WittenI \WittenIII\ that the ten-dimensional $E_{8}
\times E_{8}$ Heterotic string theory coupling constant
$\lambda_{10,H}$ is given by,
\eqn
\HeteroticCouplingConstantRealtion{
	\lambda_{10,H} = R^{3/2}_{11}.
}
\noindent In addition, one finds that the ten-dimensional $E_{8}
\times E_{8}$ Heterotic string theory metric $G_{10,H}$ and
the ten-dimensional M-Theory metric $g_{10,M}$ are related
\DavisII \WittenI \WittenIII,
\eqn
\HeteroticMTheoryMetricRelation{
	G_{10,H} = R_{11} g_{10,M}.
}
\noindent Now, let us use this information to examine the appearance
of a tensionless string in M-Theory on $S^{1} / {\bf Z}_{2}$.

M-Theory on $S^{1} / {\bf Z}_{2}$, as mentioned above, is equivalent to
the $E_{8} \times E_{8}$ Heterotic string theory. The $E_{8} \times
E_{8}$ Heterotic string theory obviously supports a one-brane. In
the M-Theory picture this one-brane arises from a M-Theory
two-brane wrapping about $S^{1} / {\bf Z}_{2}$.  Hence, if we denote
the M-Theory two-brane tension as $T_{2,M}$, then the
tension of the one-brane in the M-Theory picture $T_{1,M}$ is
given by \WittenII,
\eqn
\OneBraneTensionMTheoryMetricI{
	T_{1,M} = T_{2,M} R_{11}.
}
\noindent So, if we consider the limit in which the one-brane
tension, as measured in the M-Theory metric, goes to zero, then
we see that in this limit $T_{2,M} R_{11} \rightarrow 0$. Now, in
this case, as opposed to the Type IIA example of last section, we
have a bit of freedom in interpreting this limit. The ${\bf Z}_{2}$
factor projects out the three-form of M-Theory \WittenIII. Hence,
it also projects out the M-Theory two-branes \DuffI. Thus, the
quantity $T_{2,M}$ is not the tension of a two-brane which exists
in M-Theory on $S^{1} / { \bf Z}_{2}$, but it is the tension of a
two-brane which exists in the theory before the ${\bf Z}_{2}$
projection. So, to obtain the limit $T_{1,M} \rightarrow 0$ or
equivalently $T_{2,M} R_{11} \rightarrow 0$, we can either
take $R_{11} \rightarrow 0$ or $T_{2,M} \rightarrow 0$. Taking
$T_{2,M} \rightarrow 0$, in contrast to the Type IIA
case, does not correspond to taking the tension of a two-brane
in M-Theory on $S^{1} / {\bf Z}_{2}$ to zero tension. It corresponds
only to taking an ``internal parameter" of the theory to zero.
So, we are free to take $T_{2,M} \rightarrow 0$ or $R_{11}
\rightarrow 0$ in both cases we only obtain a tensionless
one-brane and no two-brane. Also, let us
require that the M-Theory ten-metric $g_{10,M}$ be
invariant in either limit, $g_{10,M} \rightarrow g_{10,M}$.
In the limit $R_{11} \rightarrow 0$ the spacetime, from a
M-Theory perspective, takes the form presented in Figure 3A.

\centerline{
	\epsfbox{ 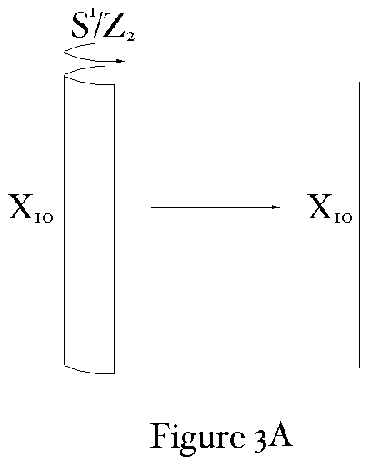 }
}

\noindent Again, $X_{10}$ is an arbitrary ten-manifold.
However, if we consider the limit in which
$T_{2,M} \rightarrow 0$, then the eleven-manifold does
not become singular. In particular, $g_{10,M}$ and
$R_{11}$ are invariant in this limit. Let us now look at
both of these limits from an $E_{8} \times E_{8}$
Heterotic string theory perspective.

As mentioned in the previous section, a $p$-brane tension is a
dimensionful quantity and thus depends upon the metric in
which it is measured. Hence, the tension of the M-Theory
one-brane is not the same when measured in the $E_{8} \times
E_{8}$ Heterotic string theory metric. So, in particular,
employing the relation between the M-Theory ten-metric
and the $E_{8} \times E_{8}$ Heterotic string theory ten-metric
\HeteroticMTheoryMetricRelation\ one finds that the one-brane
tension $T_{1,H}$  as measured in the $E_{8} \times E_{8}$
Heterotic string theory metric is given by,
\eqn
\HeteroticOneBraneTension{
	T_{1,H} = R^{-1}_{11} T_{1,M} = T_{2,M},
}
\noindent where the first equality follows from
\HeteroticMTheoryMetricRelation\ and the second from
\OneBraneTensionMTheoryMetricI.

So, from \HeteroticOneBraneTension\ one can see that
in the limit $R_{11} \rightarrow 0$ the one-brane tension
in the $E_{8} \times E_{8}$ Heterotic string theory does
not vanish. However, the ten-manifold $X_{10}$ on which
the Heterotic string theory resides does become singular.
One can easily see this from equation
\HeteroticMTheoryMetricRelation\ along with the fact that
$g_{10,M} \rightarrow g_{10,M}$. Graphically, this is
represented by Figure 3B.

\centerline{
	\epsfbox{ 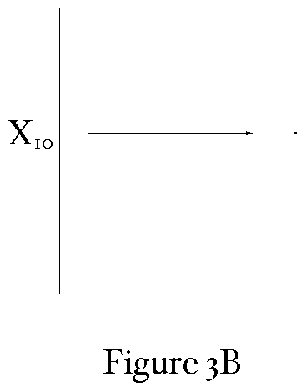 }
}

\noindent As the limit $R_{11} \rightarrow 0$ is singular from
an $E_{8} \times E_{8}$ Heterotic string theory perspective, let
consider our other option, the limit in which
$T_{2,M} \rightarrow 0$.

In this limit, according to \HeteroticOneBraneTension,
one obtains a tensionless string in the $E_{8} \times E_{8}$
Heterotic string theory. This, however, does not help our
case as we do not know how to deal with the dynamics of
such a string. But, as one can see from
\HeteroticMTheoryMetricRelation, the ten-manifold upon
which the $E_{8} \times E_{8}$ Heterotic string theory lives
does not become singular in this limit. Graphically, this limit
is represented in the same manner as depicted in Figure 2A.
However, as there exists a tensionless string in this non-singular
Heterotic theory, we do not know how to work with this
theory in this $T_{2,M} \rightarrow 0$ limit. Now, let
us consider reversing this process and looking at a tensionless
string in the $E_{8} \times E_{8}$ Heterotic string theory
from a M-Theory point-of-view.

%
%

\subsec{
Tensionless Strings in the $E_{8} \times E_{8}$ Heterotic String Theory
}

In this subsection we will examine the appearance of a tensionless
string in the $E_{8} \times E_{8}$ Heterotic string theory. We will
then interpret this string in a M-Theory context employing the
equality \WittenIII\ of M-Theory on $S^{1} / {\bf Z}_{2}$ and the
$E_{8} \times E_{8}$ Heterotic string theory. Again, as most of the
``leg-work" was performed in the previous subsection, we will
refer to it heavily in this subsection.

According to equation \HeteroticOneBraneTension\ the one-brane
tension $T_{1,H}$ as measured in the $E_{8} \times E_{8}$ Heterotic
string theory metric is $T_{1,H} = T_{2,M}$. Hence, the limit in
which the $E_{8} \times E_{8}$ Heterotic string becomes tensionless
$T_{1,H} \rightarrow 0$ corresponds to the limit $T_{2,M} \rightarrow
0$. Furthermore, if we require that $G_{10,H} \rightarrow G_{10,H}$
in this limit , then the Heterotic limit is graphically represented by
the situation in depicted Figure 2A. Now, let us look at the
implications of this tensionless string from a M-Theory perspective.

As we showed above, $T_{2,M} \rightarrow 0$ in the limit
$T_{1,H} \rightarrow 0$. However, according to
equation \OneBraneTensionMTheoryMetricI\ the one-brane
tension in M-Theory is $T_{1,M} = T_{2,M} R_{11}$. So, the
M-Theory one-brane tension not only depends upon $T_{2,M}$,
but also upon $R_{11}$. Hence, we are left with a set of choices as
to what we can take $R_{11}$ to in this limit. If we require
$T_{1,M}$ to be finite, then we must take $R_{11}$ to
infinity as $T_{2,M} \rightarrow 0$. This, while not yielding a
singular $S^{1} /  {\bf Z}_{2}$ from a M-Theory perspective
yields a singular ten-manifold. This follows from
\HeteroticMTheoryMetricRelation\ and the fact that
$G_{10,H} \rightarrow G_{10,H}$. So, as we ``generically" have
no means to deal with M-Theory on such a manifold,
let us consider some of the other possible limits.

If we allow $T_{1,M}$ to go to zero, then we can let $R_{11}$
remain constant. This does not yield a singular $S^{1} / {\bf Z}_{2}$
or ten-manifold. However, it does yield a tensionless string
in the M-Theory picture. Hence, as we do not know how to
deal with such a tensionless string, this $T_{1,M}$ limit does not
inform us in any substantial way. Finally, we may let $T_{1,M}$
go to zero while also taking $R_{11}$ to zero. Again, this limit lands
us in a theory about which little is known. As $R_{11}$ goes to
zero the $S^{1} / {\bf Z}_{2}$ factor is singular and little is known
of M-Theory  with $T_{1,M} = 0$ and $T_{2,M} = 0$ on such a
manifold. Now, let us finally look at M-Theory on $T^{2}$.
In many ways we will find this to be the most interesting
case. Also, it is the one over which we will have the most
control as we will be able to employ T-Duality to resolve
some of the singular $S^{1}$'s we will encounter.

%
%

\newsec{ Tensionless Strings : M-Theory on $T^{2}$ }

In this section we will examine the tensionless strings which
appear in M-Theory on $T^{2}$. After this we will examine
these tensionless strings from a Type II perspective.
Let us now start this examination by looking at the tensionless
strings in M-Theory on $T^{2}$.

M-Theory on $T^{2}$, as was proven by Witten \WittenI,
is equivalent to the Type IIA string theory on $S^{1}$.
The Type IIA string theory on $S^{1}$ obviously possess
a one-brane; in fact it possess two different one-branes.
It possess the standard one-brane of Type IIA
string theory, but it also possess a second one-brane which
arises from the two-brane of ten-dimensional Type IIA
string theory wrapping about $S^{1}$. From a M-Theory
perspective these two different one-branes have a similar
origin.

Type IIA string theory on $S^{1}$ is equivalent
to M-Theory on $T^{2}$ \WittenI. Hence, one could
consider the M-Theory two-brane wrapping about
either of the $S^{1}$'s which ``reside" in $T^{2} = S^{1}
\times S^{1}$. Upon wrapping the M-Theory two-brane
about the first $S^{1}$ one obtains the first one-brane.
Upon wrapping the M-Theory two-brane about the
second $S^{1}$ one obtains the second one-brane.

From a M-Theory point-of-view we can easily compute
the tension of both these one-branes. If we denote the
radius of the first $S^{1}$ as $R_{10}$ and the radius of
the second $S^{1}$ as $R_{11}$, both measured in the
M-Theory metric, then the tension of the first one-brane
$T'{}_{1,M}$ as measured in the M-Theory metric is
\eqn
\OneBraneTensionOne{
	T'{}_{1,M} = T_{2,M} R_{10},
}
\noindent where $T_{2,M}$ is the tension of the M-Theory
two-brane as measured in the M-Theory metric. Similarly,
the tension of the second one-brane $T_{1,M}$ as measured
in the M-Theory metric is
\eqn
\OneBraneTensionTwo{
	T_{1,M} = T_{2,M} R_{11}.
}
\noindent Now, one can easily see that to obtain a tensionless
one-brane one can take $R_{10} \rightarrow 0$ or $R_{11}
\rightarrow 0$. Also, one could take $T_{2,M} \rightarrow 0$
to obtain two tensionless one-branes. However, as we are
interested only in understanding the appearance of a
tensionless one-brane or one-branes in M-Theory, we
will not consider the limit $T_{2,M} \rightarrow 0$ as
it introduces, in addition to tensionless one-branes, a
tensionless two-brane. Also, let us assume that in either
of the two above limits the metric on the non-compact
nine-manifold is invariant, i.e. $g_{9,M} \rightarrow g_{9,M}$.
Let us now look at these  limits, $R_{10} \rightarrow 0$ and
$R_{11} \rightarrow 0$, from a Type II perspective.

Now, before we proceed in interpreting these tensionless
strings in terms of a Type II string theory, we must
first establish various relations between the variables of
M-Theory and those of the Type II string theory.
To some extent this was already done in section two.
The ten-metric of the Type IIA string theory $G_{10,IIA}$,
before compactification on $S^{1}$, is related to the
ten-metric $g_{10,M}$ of M-Theory as follows,
\eqn
\MetricRelationII{
	G_{10,IIA} = R_{11} g_{10,M}.
}
\noindent Similarly, the ten-dimensional Type IIA coupling
constant $\lambda_{10,IIA}$ is related to the $S^{1}$ radius
$R_{11}$ as follows,
\eqn
\TypeIIACouplingRelationII{
	\lambda_{10,IIA} = R^{3/2}_{11}.
}
\noindent Now, let us employ this information to interpret
the tensionless strings appearing in M-Theory on $T^{2}$
in terms of a Type II theory.

Consider first the limit in which $R_{11} \rightarrow 0$. As
we showed previously, the tension of the second one-brane
as measured in the M-Theory metric is $T_{1,M} = T_{2,M}
R_{11}$. So, as we are assuming $T_{2,M} \rightarrow T_{2,M}$,
in this limit $T_{1,M} \rightarrow 0$. Similarly, upon looking at
equation \OneBraneTensionOne\ one sees that
$T'{}_{1,M} \rightarrow T'{}_{1,M}$ in this limit. So, we
obtain a single tensionless string in the limit $R_{11}
\rightarrow 0$.

Now, as we are assuming that $R_{10} \rightarrow R_{10}$
and $g_{9,M} \rightarrow g_{9,M}$ in this limit, the relation
\MetricRelationII\ implies that in
the limit $R_{11} \rightarrow 0$ the ten-manifold\foot{ Again,
$X_{9}$ is an arbitrary nine-manifold.} $X_{9} \times S^{1}$
upon which the Type IIA theory is formulated is becoming
singular. In particular, the metric $G_{10,IIA}$ for this
ten-manifold is ``vanishing." Hence, the interpretation of
this particular limit from a Type IIA perspective
does not seem to teach us much about the tensionless
string in M-Theory on $T^{2}$. So, let us consider the
second limit, that in which $R_{10} \rightarrow 0$.

In this limit, as one can easily see from equations
\OneBraneTensionOne\ and
\OneBraneTensionTwo\ along with the
assumption that $T_{2,M} \rightarrow T_{2,M}$,
the tension $T'{}_{1,M}$ goes to zero and the
tension $T_{1,M}$ is constant in this limit.
Furthermore, as one can see from
\MetricRelationII, the
ten manifold $X_{9} \times S^{1}$ does not
collapse in the same manner as it did in the limit
$R_{11} \rightarrow 0$. In the case at hand, the
$S^{1}$ factor of $X_{9} \times S^{1}$ goes to zero
radius from a M-Theory perspective. Let us now
consider what happens to the $S^{1}$ radius in the
Type II picture.

Looking at \MetricRelationII\
we can see that the ten-metric of the Type IIA
theory is scaled relative to the M-Theory ten-metric.
This relation implies that the radius $R_{10}$ of the
$S^{1}$ as measured in the M-Theory metric is
different from the radius $R_{10,IIA}$ of the $S^{1}$
as measured in the Type IIA metric. In particular,
\MetricRelationII\ implies,
\eqn
\TypeIIARadius{
	R_{10,IIA} = R^{1/2}_{11} R_{10}.
}
\noindent So, in the limit we are considering, $R_{10}
\rightarrow 0$ and $R_{11} \rightarrow R_{11}$, the
radius $R_{10,IIA}$ goes to zero, as it does in the
M-Theory picture.

Now, at first this seems a little disappointing. It looks
as if in the Type IIA theory we are again on a singular
manifold as $R_{10,IIA} \rightarrow 0$. However, in
this case we may employ T-Duality to resolve this limit.
We can employ the standard T-Duality transformation
\SeibergI\  to interpret this limit in terms of a Type IIB
string theory. According to Seiberg et. al. \SeibergI\ Type
IIA string theory on a $S^{1}$ with radius $R_{10,IIA}$
is equivalent to a Type IIB string theory on a $S^{1}$
with radius $R_{10,IIB} = 1 / R_{10,IIA}$. In addition,
the coupling constants of the theories are related by
\eqn
\CouplingRelation{
	{ {R_{10,IIA}} \over {\lambda^{2}_{10,IIA}} } =
	{ {R_{10,IIB}} \over {\lambda^{2}_{10,IIB}} },
}
\noindent where $\lambda_{10,IIB}$ is the ten-dimensional
coupling constant of the Type IIB string theory.

Now, in the limit $R_{10} \rightarrow 0$, as we found
previously, $R_{10,IIA} \rightarrow 0$. This in turn
implies that $R_{10,IIB} \rightarrow \infty$. Hence,
the problem of the singular $S^{1}$ is solved if we consider
instead the Type IIB theory on $S^{1}$. In this limit
one finds that the tensionless string in M-Theory
appears just as the $S^{1}$ on which the Type IIB theory
is compactified becomes ${\bf R}$. So, we have found that
in the limit in which one of the tensionless strings appears in
M-Theory on $X_{9} \times T^{2}$ M-Theory on $X_{9}
\times T^{2}$ is equivalent to the Type IIB string theory in
ten-dimensions on $X_{9} \times {\bf R}$.

This is indeed very interesting, however, there is a bit
of a hitch. From equation \CouplingRelation,
by employing the relation between the Type IIA and Type IIB
radii along with the relation between the M-Theory radii
and the Type IIA variables, one can compute the Type
IIB coupling constant in ten-dimensions $\lambda_{10,IIB}$
in terms of the radii of M-Theory on $T^{2}$. One finds,
\eqn
\TypeIIBCouplingConstant{
	\lambda_{10,IIB} = R_{11} / R_{10}.
}
\noindent So, in the limit we are considering, $R_{10}
\rightarrow 0$ and $R_{11} \rightarrow R_{11}$, the
ten-dimensional Type IIB coupling constant goes
to infinity. At first this looks to be a bit of a problem.
But, upon a closer examination one find that this is
indeed not the case.

The Type IIB string theory in ten-dimensions
possess a $SL( 2, {\bf Z} )$ symmetry \HullI. This symmetry,
among other things, acts on the coupling constant
of the Type IIB theory. In particular \DasI, it
exchanges the weak and strong coupling regions
of the theory. Hence, we can describe our above
limit of a Type IIB string theory on $X_{9} \times
{\bf R}$ with $\lambda_{10,IIB} \rightarrow \infty$
by  a Type IIB string theory on $X_{9} \times {\bf R}$
with $\lambda_{10,IIB} \rightarrow 0$. In other
words, we have found how to describe the appearance
of a tensionless string in M-Theory on $X_{9} \times
T^{2}$ by a weakly coupled Type IIB string theory on
$X_{9} \times {\bf R}$.

However, as in all the other cases with which we have
been dealing, there is a hitch. Consider the tension
of the one-brane $T'{}_{1,M}$. In accord with our
earlier remarks, it is given by $T'{}_{1,M} = T_{2,M} R_{10}$.
Now, as the ten-dimensional Type IIA
metric $G_{10,IIA}$ and the ten-dimensional
M-Theory metric $g_{10,M}$ are related as in equation
\MetricRelationII, this implies that the tension
of this one-brane $T'{}_{1,IIA}$ as measured in
the Type IIA metric is
\eqn
\TypeIIAOneBraneTensionPrime{
	T'{}_{1,IIA} = 
	R^{-1}_{11} T'{}_{1,M} =
	R_{10} R^{-1}_{11} T_{2,M}.
}
\noindent So, in the limit $R_{10} \rightarrow 0$,
$R_{11} \rightarrow R_{11}$, and $T_{2,M} \rightarrow T_{2,M}$
the tension of this one-brane from a Type IIA perspective
goes to zero. Now, in accord with T-Duality \DasI, the
Type IIA metric $G_{9,IIA}$ on the nine-manifold $X_{9}$
is related to the Type IIB metric $G_{9,IIB}$ on the same
nine-manifold by
\eqn
\TypeIIATypeIIBMetricRelation{
	G_{9,IIA} = G_{9,IIB}.
}
\noindent Hence, in particular, any tension measured in
the Type IIA metric in nine-dimensions coincides with
the same tension as measured in the Type IIB metric. So, in
particular,
\eqn
\TypeIIBOneBraneTensionPrime{
	T'{}_{1,IIB} = T'_{1,IIA},
}
\noindent where $T'{}_{1,IIB}$ is the tension of the one-brane
$T'{}_{1,IIA}$ as measured in the Type IIB metric. Thus, as
the tension $T'{}_{1,IIA} \rightarrow 0$ in the limit we are
considering, so also $T'{}_{1,IIB} \rightarrow 0$ in the limit
we are considering. Hence, the tension of the one-brane in
the Type IIB theory on $X_{9} \times {\bf R}$ goes to zero
in the limit we are considering.

Thus, even though we are considering a weakly coupled
Type IIB string theory on $X_{9} \times {\bf R}$, we are
also considering the limit in which\foot{ Note, one may
think that the  $T'{}_{1,IIB} \rightarrow 0$ limit may be
exchanged for another limit by way of $SL( 2, {\bf Z} )$;
however, \SchwarzI\ this is not the  case. } $T'{}_{1,IIB}
\rightarrow 0$. Hence, as there is a tensionless string
in the Type IIB spectrum, we actually do not know how
to deal with this limit.

%
%

\newsec{ Generic Considerations }

We have shown that many of the limits in which a tensionless
string appears in M-Theory can be interpreted as singular
limits of String Theory or limits in which a tensionless string
appears in String Theory. However, one may also wonder if this
suggests some general trend. One can relatively easily see that
this is indeed the case.

Consider M-Theory compactified down to $d$ dimensions
on some manifold $K$. Let the $d$ dimensional M-Theory 
metric be denoted by $g_{d,M}$. Also, let us assume that
M-Theory compactified down to $d$ dimensions on $K$ is
equivalent to some string theory $S$ let us call it. Let us
denote the $d$ dimensional string theory metric as $G_{d,S}$.
Furthermore, let us assume that the two metrics are related
by some scaling factor $Q$,
\eqn
\GenericMetricRelation{
	G_{d,S} = Q g_{d,M}.
}
\noindent Now, assume that there exists a one-brane in the
M-Theory spectrum and its tension is denoted by $T_{1,M}$.
Thus, the tension $T_{1,S}$ of this one-brane as measured in
the string theory $S$ is,
\eqn
\GenericOneBraneTension{
	T_{1,S} = Q^{-1} T_{1,M}.
}

Now, let us consider the limit in which $T_{1,M} \rightarrow 0$
and $g_{d,M} \rightarrow g_{d,M}$. In this limit, if we wish to
understand it from the perspective of the string theory $S$,
we must require $T_{1,S} \rightarrow T_{1,S}$. So, this
in turn implies $Q \rightarrow 0$. Now, as $g_{d,M}
\rightarrow g_{d,M}$ and $Q \rightarrow 0$, we find
that, in accord with \GenericMetricRelation, $G_{d,S}
\rightarrow 0$. Hence, from a string theory perspective,
if we wish to maintain a string of finite tension, we must
have a singular $d$ dimensional spacetime. 

In addition, one could consider, instead of the limit in
which $T_{1,S} \rightarrow T_{1,S}$, the limit in which we
require $G_{d,S} \rightarrow G_{d,S}$, $g_{d,M} \rightarrow
g_{d,M}$, and $T_{1,M} \rightarrow 0$. As $G_{d,S}
\rightarrow G_{d,S}$ and $g_{d,M} \rightarrow g_{d,M}$ this
implies, in accord with \GenericMetricRelation, that
$Q \rightarrow Q$. So, in accord with
\GenericOneBraneTension, this implies that $T_{1,S}
\rightarrow 0$ as a result of $T_{1,M} \rightarrow 0$. So,
in this case we do not obtain a singular $d$ dimensional
spacetime from a string theory perspective, but we do
obtain a tensionless string with which we do not know
how to work.

So, it seems this is a general phenomena. The appearance
of a tensionless string in M-Theory on a non-singular
spacetime can be interpreted as the appearance of a
tensionless string in string theory on a non-singular
spacetime. Or it can be interpreted as the collapse
of spacetime with a string of finite tension from a
string theory perspective.

\listrefs
\bye